\documentstyle[11pt,amssymb,epsf,cite]{article}

\makeatletter
\@addtoreset{equation}{section}
\makeatother

\def\ben{\begin{equation}}
\def\een{\end{equation}}

\let\a=\alpha    \let\e=\epsilon
    
  \let\n=\nu   
\let\s=\sigma   \let\f=\phi  

\let\C=\Chi

\def\nn{\nonumber} \def\bd{\begin{document}} \def\ed{\end{document}}
\def\ds{\documentstyle} \let\fr=\frac \let\bl=\bigl \let\br=\bigr
\let\Br=\Bigr \let\Bl=\Bigl
\let\bm=\bibitem
\let\na=\nabla
\let\pa=\partial \let\ov=\overline
\newcommand{\be}{\begin{equation}}
\newcommand{\ee}{\end{equation}}
\def\ba{\begin{array}}
\def\ea{\end{array}}
\def\ft#1#2{{\textstyle{{\scriptstyle #1}\over {\scriptstyle #2}}}}
\def\fft#1#2{{#1 \over #2}}
\def\del{\partial}
\def\vp{\varphi}
\def\sst#1{{\scriptscriptstyle #1}}
\def\oneone{\rlap 1\mkern4mu{\rm l}}
\def\td{\tilde}
\def\wtd{\widetilde}
\def\ie{\rm i.e.\ }
\def\dalemb#1#2{{\vbox{\hrule height .#2pt
        \hbox{\vrule width.#2pt height#1pt \kern#1pt
                \vrule width.#2pt}
        \hrule height.#2pt}}}
\def\square{\mathord{\dalemb{6.8}{7}\hbox{\hskip1pt}}}
\newcommand{\ho}[1]{$\, ^{#1}$}
\newcommand{\hoch}[1]{$\, ^{#1}$}
\newcommand{\bea}{\begin{eqnarray}}
\newcommand{\eea}{\end{eqnarray}}
\newcommand{\ra}{\rightarrow}
\newcommand{\lra}{\longrightarrow}
\newcommand{\Lra}{\Leftrightarrow}
\newcommand{\ap}{\alpha^\prime}
\newcommand{\bp}{\tilde \beta^\prime}
\newcommand{\tr}{{\rm tr} }
\newcommand{\Tr}{{\rm Tr} }
\def\0{{\sst{(0)}}}
\def\1{{\sst{(1)}}}
\def\2{{\sst{(2)}}}
\def\3{{\sst{(3)}}}
\def\4{{\sst{(4)}}}
\def\5{{\sst{(5)}}}
\def\6{{\sst{(6)}}}
\def\7{{\sst{(7)}}}
\def\8{{\sst{(8)}}}
\def\n{{\sst{(n)}}}
\def\cA{{{\cal A}}}
\def\cF{{{\cal F}}}
\def\tV{\widetilde V}
\def\tW{\widetilde W}
\def\tH{\widetilde H}
\def\tE{\widetilde E}
\def\tF{\widetilde F}
\def\tA{\widetilde A}
\def\im{{{\rm i}}}
\def\tY{{{\wtd Y}}}
\def\ep{{\epsilon}}
\def\vep{{\varepsilon}}
\def\R{\rlap{\rm I}\mkern3mu{\rm R}}
\def\bD{{{\bar D}}}

\def\R{\rlap{\rm I}\mkern3mu{\rm R}}
\def\bD{{{\bar D}}}
\def\R{{{\Bbb R}}}
\def\C{{{\Bbb C}}}
\def\H{{{\Bbb H}}}
\def\CP{{{\Bbb C}{\Bbb P}}}
\def\RP{{{\Bbb R}{\Bbb P}}}
\def\Z{{{\Bbb Z}}}
\def\bA{{{\Bbb A}}}
\def\bB{{{\Bbb B}}}
\def\bC{{{\Bbb C}}}
\def\bR{{{\Bbb R}}}
\def\bD{{{\Bbb D}}}
\def\bE{{{\Bbb E}}}
\def\bZ{{{\Bbb Z}}}
\def\Re{{{\frak{Re}}}}
\def\Im{{{\frak{Im}}}}
\def\cosec{{\,\hbox{cosec}\,}}
\def\Gm{{\Gamma_{\!\! -}}}
\def\Gp{{\Gamma_{\!\! +}}}
\def\stan{{standard }}
\def\nonstan{{supernumerary }}

\def\cosech{{\hbox{cosech}}}

\def\etcyc{{\hbox{and cyclic}}}
\def\btheta{{\bar\theta}}

\def\vf{{\varphi}}
\def\hf{{\hat\phi}}
\def\fh{{\hat f}}
\def\ah{{\hat a}}
\def\eq#1{(\ref{#1})}
\newcommand{\w}[1]{\\[0.#1cm]}
\def\eb{ {\bar\epsilon} }
\def\cA{{\cal A}}
\def\cB{{\cal B}}
\def\cF{{\cal F}}
\def\cV{{\cal V}}
\def\cG{{\cal G}}
\def\cP{{\cal P}}
\def\cQ{{\cal Q}}
\def\hL{{\hat L}}
\def\ua{{\underline{\alpha}}}
\def\ub{{\underline{\beta}}}
\def\uc{{\underline{\gamma}}}

\newcommand{\undertilde}[1]{\underset{\widetilde{}}{#1}}

\newcommand{\tamphys}{\it George P. and Cynthia W. Mitchell  Institute
for Fundamental Physics,\\
Texas A\&M University, College Station, TX 77843, USA}

\newcommand{\auth}{
H. L\"u, C.N. Pope and E. Sezgin
}

\begin{document}
\begin{flushright}
\hfill{
MIFP-06-37}\\
\hfill{
\bf hep-th/0612293}
\end{flushright}

\begin{center}

{\large {\bf Group Reduction of Heterotic Supergravity
}}

\vspace{15pt}

\auth

\vspace{7pt}
{\hoch{\ddagger}\tamphys}

\vspace{15pt}

\underline{ABSTRACT}

\end{center}

\vspace{15pt}

The reduction of ten-dimensional heterotic supergravity with
Yang-Mills symmetry group $K$ is performed on an arbitrary
$n$-dimensional group manifold $G$. The reduction involves a
nonvanishing $3$-form flux, and the Lie algebra of $G$ must have
traceless structure constants to ensure the consistency of the
reduction at the level of the action. A large class of gauged
supergravities in $d=10-n$ with (non)compact gaugings is obtained.
The resulting models describe half-maximal gauged supergravities
coupled to $ (n + {\rm dim}\,K)$ vector multiplets. We uncover their
hidden $SO(n,n+{\rm dim}\,K)$ duality symmetry, and the $SO(n,n+{\rm
dim}\,K) / SO(n)\times SO(n+{\rm dim}\,K)$ coset structure that
governs the couplings of the scalar fields. We find that the local
gauge symmetry of the $d$-dimensional theory is $K\times G \ltimes
R^n$. Differences from the existing gauged supergravities are
highlighted. The consistent truncation to pure half-maximal gauged
supergravity in any dimension is shown, and the obstacle to
performing a chiral truncation of the theory in $d=6$ dimensions is
found. Among the results obtained are the complete diagonalisation
of the fermionic kinetic terms, and other reduction formulae that
are applicable to group reductions of supergravities in arbitrary
dimensions.

\pagebreak

\setcounter{page}{1}

\tableofcontents

\addtocontents{toc}{\protect\setcounter{tocdepth}{2}}

\newpage


\newpage


\section{Introduction}


In this paper, we study the group manifold ``DeWitt reduction'' of
ten-dimensional heterotic supergravity. More specifically, we study
the reduction of the theory on a group manifold $G$ of dimension
$n$. The group need not be compact or simple, but the structure
constants of the underlying Lie algebra must be traceless to ensure
consistent reduction at the level of the action
\cite{Scherk:1979zr}.

This work grew out of our attempts to understand the string/M-theory
origin of the remarkable anomaly-free gauged supergravities that exist in six
dimensions \cite{Nishino:1984gk,Salam:1984cj}. Models of this type
have been increasingly finding applications in cosmology, and in
braneworld scenarios
\cite{Halliwell:1986bs,Maeda:1985es,Aghababaie:2003wz,Gibbons:2003di,
Nair:2004yu,Carter:2006uk}. One of the anomaly-free models, which has
$E_7\times E_6\times U(1)_R$ symmetry, was found long ago
\cite{Randjbar-Daemi:1985wc}. More recently anomaly-free models with
$E_7\times G_2\times U(1)_R$  symmetry \cite{Avramis:2005qt} and
$F_4\times Sp(9)\times U(1)_R$ symmetry \cite{Avramis:2005hc} have
been found.

Progress has been made in embedding a minimal sub-sector with
$U(1)_R$ symmetry and no hyperfermions in string/M theory
\cite{Cvetic:2003xr}. This result has been generalized to some
extent, to include a larger set of fields, in
\cite{Bergshoeff:2005pq}. In both of these efforts a key role is
played by noncompact gaugings in half-maximal supergravities in
$d=7$  coupled to a number of vector multiplets. In particular, an
$SO(2,2)$ gauged theory is reduced on a circle to $d=6$ and then
chirally truncated to obtain the desired result. In
\cite{Cvetic:2003xr} it was argued that the theory in $d=7$  can itself be
obtained via a consistent reduction of heterotic supergravity on a certain
3-manifold $H_{2,2}$, which is a hyperboloid embedded in $R^4$.
In \cite{Bergshoeff:2005pq}, on the other
hand, certain noncompact gauged theories were directly considered in
$d=7$, and their chiral reduction to $d=6$ was obtained. Among these
models were an $SO(2,1)$ gauged half-maximal supergravity coupled to
a single vector multiplet. Given that gauged supergravities are
known to result from a DeWitt (or Scherk-Schwarz) reduction on a group
manifold,
it became natural to consider the prospects of obtaining the
$SO(2,1)$ gauged model from the group reduction of the heterotic
supergravity theory. This has been the prime motivation for the
results which will be reported in this paper.

To begin with, we find the full bosonic Lagrangian in $d$ dimensions
that results from the reduction on the group manifold $G$.
It describes the coupling of half-maximal supergravity to $(n+{\rm
dim}\, K)$ vector multiplets. The reduced theory is invariant under
local $K\times G\ltimes R^n$ transformations. Here, $G$ is any
$n$-dimensional group based on a Lie algebra with traceless
structure constants.  Such algebras are sometimes referred to as
type A, to distinguish them from
those whose structure constants have a non-vanishing trace,
which are referred to as
type B. It is necessary to work with type A Lie algebras in order to ensure
that the group reduction is consistent when the ansatz is substituted into
the higher-dimensional action, as was
observed in \cite{Scherk:1979zr}.   By contrast, a type B reduction works
consistently only at the level of the field equations.  Although substitution
of the reduction ansatz for a type B algebra into the higher-dimensional
equations of motion consistently yields equations of motion for the
lower-dimensional theory, these equations cannot themselves
be derived from an action.  This was
discovered first in \cite{hawking}, in the study of homogeneous cosmological
models, and was more recently utilized in
\cite{eric} in the group reduction of maximal supergravities.

A complete list of Lie algebras up to dimension five has been given
in \cite{4dLie},  although we shall use the notation of the list provided
in \cite{Patera:1976ud}. There are 12 three-dimensional algebras, 20
four-dimensional algebras and 40 five-dimensional algebras that are not
themselves direct
sums of lower-dimensional algebras.  Of these, there are 5, 4 and 19
respectively that are isolated type A Lie algebras.  In addition,
there are 2 examples in dimension four, and 4 examples in dimension five,
that have a non-trivial free parameter, and a further 3 examples in dimension
five that are two-parameter families of Lie algebras.
These algebras with non-trivial parameters introduce new arbitrary parameters
in the gauged supergravities obtained by the group reduction.

We also perform the group reduction of the supersymmetry
transformations, up to cubic order in fermions. In doing so, we  derive
general formulae that can be used in the group reductions of a large
class of supergravity theories in diverse dimensions.  In
particular, we obtain a complete diagonalisation of all the fermion
kinetic terms.  This goes beyond what has generally been done in the
earlier literature.

Next, we exhibit the hidden $SO(n,n+{\rm dim}\,K)$ duality symmetry of
the $d$-dimensional theory, and the $SO(n,n+{\rm dim}\,K) /
SO(n)\times SO({\rm dim}\,K)$ coset structure that governs the
couplings of the scalar fields. We do so by first setting the gauge
couplings to zero, thereby reducing the models to half-maximal supergravity
coupled to $(n+{\rm}\, K)$ Maxwell multiplets.  This symmetry structure has
already been exhibited in \cite{Maharana:1992my} at the level of the
bosonic action, in the context of toroidal reduction. Here we
generalize those results to include the supersymmetry transformation
rules. We then generalize further by turning on the
gauge couplings, finding dramatically simplified formulae for the
action and supersymmetry transformation rules in an arbitrary dimension
$d$. In particular, the potential takes a simple and universal form
with a transparent  group-theoretical structure. Thus, we have
embedded a large class of gauge theories in heterotic string theory.

Turning to the issue of the consistent truncation of these theories,
we have found that $(n+ {\rm}\,K)$ vector multiplets can always be
truncated consistently to yield a pure half-maximal gauged
supergravity theory in $d=10-n$ dimensions, in cases where the
$n$-dimensional gauge group is compact, gauging the $R$ symmetry.

By half-maximal supergravities we mean those with $16$ real
supersymmetries, and gauging refers to the $R$-symmetry group. Such
supergravities exist in $d\le 8$ and they have been constructed
directly in a given dimension over the years, and there are
scattered results on particular group manifold reductions to obtain
a small subset of them. Half-maximal gauged supergravities in
dimensions $3\le d\le 8$ can be found in
\cite{Salam:1985ns,Townsend:1983kk,Salam:1983fa,Bergshoeff:1985mr,Romans:1985tw,
Andrianopoli:2001rs,Dall'Agata:2001vb,Romans:1985ps,Awada:1985ep,Bergshoeff:1985ms,
deRoo:1985jh,Schon:2006kz,Nicolai:2001ac} and there exists some
results on group manifold reductions that yields a small set of such
theories
\cite{eric,Chamseddine:1999uy,Chamseddine:1997mc,deRoo:2005be,Bergshoeff:1997mg}.

Finally, we have examined the question that originally motivated our
work, namely the prospects for obtaining either the $SO(2,1)$ gauged
theory of \cite{Bergshoeff:2005pq}, or a direct chiral truncation in
$d=6$ with surviving gauged $R$-symmetry. We find that neither the
$SO(2,1)$ model nor the $R$-symmetry gauged chiral $d=6$ model can
be obtained from the models we have obtained by the group reduction
of the heterotic supergravity, even  if the group manifold $G$ is
taken to be $SO(2,1)$. The obstruction in $d=7$ is due to the fact
that the gauge symmetry of the model we have obtained is $K\times
SO(2,1)\ltimes R^3$, and the truncation of two vector multiplets
that is needed in order to obtain the $SO(2,1)$ gauged model of
\cite{Bergshoeff:2005pq} is not possible since they are charged
under the $SO(2,1)$ symmetry. The nonexistence of the chiral
truncation in $d=6$ is also closely related to this fact, and it is
explained in detail in Section 5.2.

The paper is organized as follows. In the next section, the
ten-dimensional heterotic supergravity model is presented in our
notation and conventions, followed by the group reduction ansatz for
all the fields. The Scherk-Schwarz formulae are generalized by
turning on the $3$-form flux. In Section 3, the full bosonic
Lagrangian and supersymmetry transformation rules, up to cubic
fermions, are obtained. The hidden symmetries are uncovered in
Section 4, and the consistent truncations, or possible obstructions
to them, are described in Section 5.  Further comments on our
results are collected in the conclusions.  An appendix contains a table 
listing of the relevant Lie algebras considered in this paper, of
dimensions 2, 3 and 4.


\section{Ten-Dimensional Heterotic Supergravity}


The Lagrangian, up to quartic fermion terms, is given by
\cite{Bergshoeff:1981um,Chapline:1982ww}
\bea
{\cal L}&=& {\cal L}_B+{\cal L}_F \w2
{\cal L}_B&=& R {*\oneone} -\ft12  {*d\f} \wedge d\f
-\ft12 e^{\ah\f} {*G} \wedge G -\ft12 e^{\ah\f/2} \, \tr' {*F}\wedge F
\nn\w2
e^{-1}{\cal L}_F &=& -{\bar\psi}_M \Gamma^{MNP} D_N\psi_P -
\ft12 {\bar\chi}\Gamma^M D_M
\chi -\ft12 \tr' {\bar\lambda}\Gamma^M D_M \lambda -\ft12 {\bar\psi}_M
\Gamma^N \Gamma^M \chi \partial\f  \nn\w2
&& +\ft1{4} e^{\ah\f/2}G_{PQR} \left({\bar\psi}^M
\Gamma_{[M}\Gamma^{PQR}\Gamma_{N]}
\psi^N -{\bar\psi}_M \Gamma^{PQR}\Gamma^M \chi +
\ft12\tr' {\bar\lambda}\Gamma^{PQR}\lambda \right)\nn\w2
&& -\ft12 e^{\ah \f/4} \tr' F_{MN}\left( {\bar\lambda}\Gamma^P
\Gamma^{MN}\psi_P
+\ft12 {\bar\lambda} \Gamma^{MN}\chi \right)\ ,
\eea
where the field strengths are defined as
\be
G = dB- \ft12 \omega_{3Y}\ , \qquad \omega_{3Y}= \tr' (FA-\ft13
A^3)\ ,\qquad   F=dA+A^2\ , \qquad D =d+[A, \ ]\ .
\ee
Note that the $\chi^2 G$ type couplings which arise in dimensions
$D<10$ are absent in ten dimensions.  In $D=10$, we have
\be \ah=-1\ ,\ee
but we shall define it as
\be \ah^2= {8\over D-2}\ ,\ee
since then the reduction formula we obtain may be used when starting from
an arbitrary dimension $D$.
We have also used the notation $A= A^I T_I$, $A^2=A\wedge A$,
$FA= F\wedge A$, etc.  The generators
are anti-hermitian, obeying the Lie algebra $[T_I,T_J]=f^K{}_{IJ} T_J$,
and normalized {\it in\ the\ fundamental\ representation} as
\be  \tr\, T_I T_J = \beta'\, \delta_{IJ}\ , \quad\quad
\beta'=\cases{ -1 \ \ \ \mbox{for}\ \  SO(32)\cr -30 \ \ \mbox{for}\
\ E_8^i\ , \ \ i=1,2} \ee
We have also defined the normalized trace in the fundamental representation as
\be \tr' = \ft1{\beta'}\, \tr\ .\ee
Note that, both for $SO(32)$ as well as $E_8$, we have $ \tr'\, T_I T_J =\delta_{IJ}$.

The supersymmetry transformations, up to terms cubic in fermion, are:
\bea
\delta e_M^a &=& \ft12 {\bar \epsilon}\Gamma^a\psi_M \nn\w2
\delta\f &=& -\ft12 {\bar\epsilon}\chi \nn\w2
\delta B_{MN} &=&
e^{-\ah\f/2}\left({\bar\epsilon}\Gamma_{[M}\psi_{N]} +\ft{\ah}{4}\,
{\bar\epsilon}\Gamma_{MN}\chi\right) - \tr' A_{[M} \delta A_{N]}
\nn\w2
\delta A_M &=& \ft12 e^{-\ah\f/4} {\bar\epsilon} \Gamma_M\lambda \nn\w2
\delta\psi_M &=& D_M\epsilon +\ft{\ah^2}{96}\, e^{\ah\f/2}
\left(\Gamma_M\Gamma^{PQR} -12\delta_M^P \Gamma^{QR}\right)
G_{PQR}\epsilon \nn\w2
\delta\chi &=& -\ft12 \Gamma^M\partial_M\f \epsilon +\ft1{24}
e^{\ah\f/2} \Gamma^{MNP} G_{MNP} \epsilon \nn\w2
\delta\lambda &=& -\ft12 e^{\ah\f/4} \Gamma^{MN} F_{MN}\epsilon
\eea
For completeness, we also record the Green-Schwarz anomaly counterterm
and the modified
Bianchi identities. The former can be read of from the anomaly
polynomial \cite{Green:1984sg}
\be \Omega_{12} = -\fft1{48\times (2\pi)^6}\, X_4 X_8\ , \ee
where
\bea
    X_4 &=& \tr F^2 -\tr R^2\ ,\nn\\
    X_8 &=&  \tr F^4-\ft18 \tr F^2\tr R^2 +
      \ft1{32} (\tr R^2)^2 + \ft18 \tr R^4 \ ,
\eea
for $SO(32)$ and
\bea X_4 &=& \ft1{30} \tr F_1^2 +\ft1{30} \tr F_2^2 -\tr R^2\ ,\nn\\
X_8&=& \ft1{3600}\left[ (\tr F_1^2)^2+ (\tr F_2^2)^2-\tr F_1^2 \tr
F_2^2\right] \nn\\
&& -\ft1{240} \left[\tr F_1^2 +\tr F_2^2\right] \tr R^2 +\ft1{32}
(\tr R^2)^2 + \ft18 \tr R^4 \ , \eea
for $E_8\times E_8$. We use the notation:
$R=\ft12 dx^M dx^N R_{MN}{}^{ab} (\ft12 T_{ab})$
with the anti-hermitian generators
$T_{ab}^{cd}= \delta_a^c\delta_b^d-\delta_a^d\delta_b^c$.

The anomalies can be  read off from $\Omega_{12}$
via the descent equations, and they are cancelled by adding the Green-Schwarz
counterterm
\be  {\cal L}_{GS}= \fft1{48 \times (2\pi)^6}\left (
       B X_8 + \ft23  X_3^0 X_7^0\right)\ ,
\ee
where $X_3^0$ and $X_7^0$ are defined by $X_4 = d X_3^0$ and
$X_8=dX_7^0$, respectively, together with a modification the $3$-form
field strength in which the Lorentz Chern-Simons term is added:
\be G \quad \rightarrow\quad G + \omega_{3L}\ .
\ee
In what follows, we shall not reduce the Green-Schwarz counterterm
to lower dimensions, since the reduced theories we shall obtain will
not have any perturbative anomalies, owing to the fact that
they are either in odd dimensions, where there are no chiral
fermions, or else in even dimensions but always vector-like. Should
any chiral truncation be possible in an even lower dimension, the
anomaly counterterms might play a role and they would have to be
considered.  Reduction of the anomaly counterterms would also make
sense if all the Kaluza-Klein modes were to be kept, which amounts to
re-writing the original theory in the lower-dimensional language. As for
the Lorentz Chern-Simons modifications of the transformation rules,
we shall not reduce them either, since they represent $\a'$
corrections to the self-contained lowest-order heterotic
supergravity.


\subsection{The reduction ansatz}


In this section we shall use to a large extent the notation and some of
the results of \cite{Cvetic:2003jy}.

\underline{{\it The Metric}:}
\medskip


We begin with the reduction ansatz for the metric:
\be d\hat s^2 = e^{2\a\varphi}\, ds^2 + e^{2\beta\varphi}\,
h_{\a\beta}\, \nu^\a\,\nu^\beta\ ,\label{ma} \ee
where
\be \nu^\a\equiv \sigma^\a - \cA^\a\ .
\ee
Here $ds^2$ is the metric in $d$ dimensions, $A^\a$ are the
Yang-Mills potentials for the gauge group $G$, and $h_{\a\beta}$ is
a unimodular symmetric matrix parameterising the scalar degrees of
freedom. The group manifold is $n$-dimensional, and thus we have
\be D=d+n\ .\ee
The Yang-Mills field strengths are
given by
\be F^\a = d\cA^\a + \ft12 f^\a{}_{\beta\gamma}\, \cA^\beta\wedge
A^\gamma\ , \label{mfs}\ee
and the constants $\a$ and $\beta$ are chosen to be
\be \a=- \sqrt {n\over 2(d-2)(D-2)}\ ,\qquad \beta=-{\a(n-2)\over
n}\ . \ee
These choices ensure that the reduction of the Einstein-Hilbert
action from $D$ dimensions yields a pure Einstein-Hilbert term, and
that $\varphi$ has a canonically-normalised kinetic term, in $d$
dimensions.

   We shall choose the vielbein basis to be
\be
{\widehat e}^a = e^{\a\varphi}\, e^a\ , \qquad {\widehat e}^i =
e^{\beta\varphi}\, L^i_\a\, \nu^\a\,,
\ee
where
\be
h_{\alpha\beta}= L^i_\alpha L^i_\beta\,.
\ee
Noting that $d {\widehat e}^A = -{\widehat\omega}^A{}_B\wedge
{\widehat e}^B$ with ${\widehat\omega}_{AB}=-{\widehat\omega}_{BA}$,
one finds \cite{Scherk:1979zr}
\bea {\widehat \omega}_{ab} &=& \omega_{ab} +\a\, e^{-\a \vf}\,
(\del_b\vf\, \eta_{ac} - \del_a\vf\, \eta_{bc}\,
)\,\hat e^c + \ft12 e^{(\beta-2\a)\, \vf}\, F^i_{ab}\, \hat e^i\ ,\nn\\
{\widehat \omega}_{ai} &=& -e^{-\a\vf}\, P_{a\, ij}\,  \hat e^j -
\beta\, e^{-\a\vf}\, \del_a\vf\, \hat e^i + \ft12 e^{(\beta-2\a)\,
\vf}\,
  F^i_{ab}\, \hat e^b\ ,\label{spincon}\\
{\widehat\omega}_{ij} &=& e^{-\a\vf}\, Q_{a\, ij}\, \hat e^a +
\ft12\, e^{-\beta\vf}\,C_{k,\,ij}\, \hat e^k\ , \eea
where $F^i=F^\a L_\a^i$ and
\be
C_{k,\,ij} = \left(L_{\a k}\, L^\beta_i\, L^\gamma_j + L_{\a
j}\, L^\beta_i\, L^\gamma_k -L_{\a i}\, L^\beta_j\, L^\gamma_k
\right)\,f^\a{}_{\beta\gamma} \,.
\ee
Using these formulae, one finds
\bea
\int_{M_{10}}  {\widehat R}\,\,  {\hat *\oneone}
&=& \fft1{n!} \int_G d^n y\,
\epsilon_{\a_1\cdots \a_n} \nu^{\a_1}
\wedge \cdots\wedge \nu^{\a_n} \int_{M_d} 
{\widehat R}_{(d)}\, {*\oneone} \nn\w2
&=& \fft1{n!}\int_G d^n y\, \epsilon_{\a_1\cdots\a_n}
\sigma^{\a_1} \wedge \cdots \wedge  \sigma^{\a_n}
\int_{M_d} {\widehat R}_{(d)}\, {*\oneone} \nn\w2
&=& \mbox{vol}(G) \int_{M_d} {\widehat R}_{(d)}\, {*\oneone} \ ,
\eea
where $\mbox{vol}(G)$ is the volume of the group manifold, and
\bea
{\widehat R}_{(7)} {*\oneone} &=&
R\, {*\oneone} - \ft12 {*d\varphi}\wedge d\varphi -
{*P_{ij}}\wedge P_{ij} - \ft12 e^{2(\beta-\a)  \varphi}\,
h_{\a\beta}\, {*\cF^\a}\wedge \cF^\beta
\nn\\
&& - \ft14 \,  e^{2(\a-\beta)\varphi}\, (h_{\a\beta}\, h^{\gamma\delta}\,
h^{\rho\sigma}\, f^\a{}_{\gamma\rho}\, f^\beta{}_{\delta\sigma} + 2
h^{\a\beta}\, f^\gamma{}_{\delta\a}\, f^\delta{}_{\gamma\beta})\, \,
{*\oneone}\ . \label{e7}
\eea
Here, we have used the notation:
\be
P_{a\, ij} \equiv  L^\a_{(i}\, D_a L_{j)\a}\ ,
\qquad
Q_{a\, ij} \equiv  L^\a_{[i}\, D_a L_{j]\a}\ ,
\ee
where $L^\a_i$ is the inverse of $L_\a^i$, and our
symmetrizations and antisymmetrizations are always with unit strength.
It is important to note that no raising or lowering of the group indices
has been performed in obtaining \eq{e7}, and so this expression is valid
for all the type A groups under consideration, even in cases where
the Cartan-Killing metric may be degenerate.

\bigskip

\underline{{\it The Yang-Mills Fields}:}

\medskip

Next, we consider the ansatz for the Yang-Mills fields:
\be {\widehat A}^I= A^I+ \phi_\a^I\,\nu^\a \,. \label{ah}
\ee
Defining the lower-dimensional components as
\be {\widehat F}^I= \left(F^I -\cF^\a \phi_\a^I\right) +P_\a^I \wedge \nu^\a
+ \ft12 F^I_{\a\beta}\,\nu^\a\wedge\nu^\beta\ ,\label{fh}\ee
one finds that
\bea
P^I_\a &=& D \phi_\a^I\ ,\nn\\
F^I_{\a\beta} &=& f^I{}_{JK}\, \phi_\a^J \phi^K_\beta
- f^\gamma{}_{\a\beta}\,\phi^I_\gamma\ ,\eea
where
\bea
F^I &=&dA^I+\ft12 f^I{}_{JK} A^J\wedge A^K\ ,\label{efs}\\
D \phi_\a^I &=& d\phi_\a^I + f^\gamma{}_{\a\beta}\cA^\beta  \phi_\gamma^I
+ f^I{}_{JK}\,A^J \phi^K_\a\ .\eea
As the ten-dimensional theory involves Yang-Mills Chern-Simons forms,
it is also useful to consider their dimensional reduction.  We define the
terms in the lower dimension by setting
\bea
{\widehat\omega}_{3Y} &=&{\widehat F}^I\wedge {\widehat A}^I
-\ft16\,f_{IJK}\,{\widehat A}^I\wedge {\widehat A}^J\wedge {\widehat A}^K\ ,\nn\\
&=& \omega'_{3Y}+ \omega_{(2)\a}\,\nu^\a +
\ft12 \omega_{(1)\a\beta}\,\nu^\a\wedge\nu^\beta +
\ft16 \omega_{(0)\a\beta\gamma}\,\nu^\a\wedge\nu^\beta\wedge\nu^\gamma
\,.
\eea
 From \eq{ah} and \eq{fh} it then follows that
\bea \omega'_{3Y} &=& \omega_{3Y}-\phi^I_\a F^\a\wedge A^I
 ,\nn\\
\omega_{(2)\a}&=& \left(F^I-\cF^\beta \phi_\beta^I\right)\phi_\a^I
 -P^I_\a\wedge A^I - \ft12 f_{IJK}A^I\wedge A^J \phi^K_\a\ ,\nn\\
\omega_{(1)\a\beta}&=& 2P^I_{[\a}\,\phi^I_{\beta]} +
F^I_{\a\beta} A^I- f_{IJK}\,A^I\phi^J_\a\phi^K_\beta\ ,\nn\\
\omega_{(0)\a\beta\gamma}&=& 3 F^I_{[\a\beta}
\phi^I_{\gamma]}-f_{IJK} \phi^I_\a \phi^J_\beta\phi^K_\gamma\ .\label{cs}
\eea

\bigskip


\underline{{\it The $2$-Form Potential}:}

\medskip

Next, we consider the $2$-form potential, for which the reduction
ansatz will be
\be
{\widehat B} = m\,\omega_\2 + B + B_{\a}\wedge
\nu^\a + \ft12 B_{\a\beta}\, \nu^\a\wedge \nu^\beta\,.
\label{b2}
\ee
Here we have introduced $\omega_\2$, defined by the requirement that
\be
d\omega_\2 = \ft16 m f_{\a\beta\gamma}\, \sigma^\a\wedge \sigma^\beta\wedge
              \sigma^\gamma\,.
\ee
The lowering of the index on the structure constants of the group $G$
is performed using the
Cartan-Killing metric $\eta_{\alpha\beta}$:
\be  f_{\a \beta\gamma}= \eta_{\gamma\delta}\,f^\delta{}_{\a\beta}\ ,
\qquad \eta_{\a\beta} = -\ft12\,f^\gamma{}_{\delta\a}\,
f^\delta{}_{\gamma\beta}\ .
\ee
This is well-defined even for non-semisimple Lie algebras where
$\eta_{\alpha\beta}$ may be degenerate, or
even vanishing. Since we wish to cover these cases as well, the group
indices will never be raised with the inverse metric $\eta^{\a\beta}$,
since this may not exist.   We define lower-dimensional field
strengths by writing
\be
d\widehat B = G^{(0)} + G^{(0)}_{\a} \wedge \nu^\a +
   \ft12 G^{(0)}_{\a\beta}\wedge \nu^\a\wedge \nu^\beta
    + \ft16 G^{(0)}_{\a\beta\gamma}\, \nu^\a\wedge \nu^\beta\wedge \nu^\gamma\ .
\ee
It follows that \cite{Cvetic:2003jy}
\bea
G^{(0)} &=& dB + B_{\a}\wedge \cF^\a + \ft16 m\, f_{\a\beta\gamma}\,
\cA^\a\wedge \cA^\beta\wedge \cA^\gamma\, \nn\\
G^{(0)}_{\a} &=& D B_{\a} + B_{\a\beta}\, \cF^\beta  + \ft12 m\,
f_{\a\beta\gamma}\, \cA^\beta\wedge \cA^\gamma\ ,\nn\\
G^{(0)}_{\a\beta} &=& DB_{\a\beta} + f^\gamma{}_{\a\beta}\,
B_{\gamma} + m\, f_{\gamma\a\beta}\, \cA^\gamma\ ,\nn\\
G^{(0)}_{\a\beta\gamma} &=& -3 B_{\delta[\a}\, f^\delta{}_{\beta\gamma]}
+ m\,f_{\a\beta\gamma}\ , \label{gs}
\eea
where $DB_{\a}=dB_{\a}+f^\gamma{}_{\a\beta} \cA^\beta B_{\gamma}$. Note that for $n=3$,
the first term on the right hand side of $G_{(0)\a\beta\gamma}$ vanishes. Expanding
${\widehat G}_{(3)}$,
\be
{\widehat G}_{(3)} = G_{(3)} + G_{(2)\a} \wedge\nu^\a +\ft12 G_{(1)\a\beta}
\wedge\nu^\a \wedge \nu^\beta
+ \ft16 G_{(0)\a\beta\gamma}\,\nu^\a\wedge\nu^\beta\wedge\nu^\gamma\ ,
\ee
where the subscripts denote the form degree, we find
\bea
G_{(3)} &=& G^{(0)}-\ft12 \omega'_{3Y}\,,\nn\\
G_{(2)\a}&=& G_\a^{(0)}-\ft12 \omega_{(2)\a}\,,\nn\\
G_{(1)\a\beta} &=& G_{\a\beta}^{(0)}-\ft12 \omega_{(1)\a\beta}\,,\nn\\
G_{\a\beta\gamma}&=& G_{\a\beta\gamma}^{(0)}-\ft12
\omega_{(0)\a\beta\gamma}\ ,\label{gc}
\eea
where the $G^{(0)}$ are given in \eq{gs} and the Chern-Simons terms in \eq{cs}.

   These results suggest that one make the field redefinition
\be
C_\a \equiv B_\a+ \ft12 A^I\phi^I_\a \ .
\ee
This is also motivated, as we shall see below, by the fact that
$C_\a$ is invariant under the gauge transformations of the
Yang-Mills group $K$, while $B_\a$ is not. With this redefinition,
the results \eq{gc} take the form:
\bea
G_{(3)} &=& dB_2 -\ft12 \omega_{3Y} +C_\a \wedge \cF^\a +
\ft16 m f_{\a\beta\gamma}
\cA^\a \cA^\beta \cA^\gamma\,, \label{g3}\\
G_{(2)\a}&=& DC_\a +C_{\a\beta} \cF^\beta - F^I\phi^I_\a
+\ft12 m f_{\a\beta\gamma} \cA^\beta \wedge \cA^\gamma \,,\label{g2}\\
G_{(1)\a\beta} &=& DB_{\a\beta} + f^\gamma{}_{\a\beta} C_\gamma
- P^I_{[\a} \phi^I_{\beta]}+ m f _{\a\beta\gamma} \cA^\gamma\,,\label{g1}\\
G_{\a\beta\gamma}&=&  3 f^\delta{}_{[\a\beta}\,
C_{\gamma]\delta}-f_{IJK} \phi_\a^I\phi_\beta^J\phi_\gamma^K  + m
f_{\a\beta\gamma}\ ,\label{gijk} \eea
where we have defined
\be
C_{\a\beta}= B_{\a\beta}+\ft12 \phi_\a^I\phi_\beta^I\ .
\ee
The resulting Bianchi identities are
\bea dG_\3 &=& G_\a\wedge \cF^\a
-\ft12\left(\cF^\a\phi^I_\a-F^I\right)
\left(\cF^\beta\phi^I_\beta-F^I\right)\,,\w2
DG_{\2\a}&=& G_{\a\beta}\wedge\cF^\beta+ P_\a^I\wedge
\left(\cF^\beta\phi^I_\beta-F^I\right)\,,\w2
DG_{\1\a\beta}&=& G_{\a\beta\gamma}\,\cF^\gamma +P_\a^I\wedge P_\beta^I
\nn\w2
&&+f^\gamma{}_{\a\beta}\left(G_{\2\gamma}+F^I\phi^I_\gamma\right)
+f_{IJK}\phi^I_\a\phi^J_\beta
\left(\cF^\gamma\phi^K_\gamma-F^K\right) \,.
\eea
 Finally, we note that the gauge
transformation of the $2$-form potential,
\be
\delta {\widehat B}= \ft12 {\widehat A}^I d {\widehat\Lambda}^I\,,
\ee
implies the following $\Lambda_I$ gauge transformations in seven dimensions:
\be
\delta B =\ft12 A^I\wedge d\Lambda^I\ ,\qquad \delta C_\a=0\ ,
\qquad \delta B_{\a\beta}=0\,.
\ee
%


\section{The Model in $d$ Dimensions}


\subsection{The bosonic Lagrangian}


The reduction of the kinetic term for the $2$-form potential yields the result
\cite{Scherk:1979zr,Cvetic:2003jy}
\bea
{\cal L}_3 &=& -\ft12 e^{-(\hf+4\a\vf)}\,{*G_{(3)}} \wedge G_{(3)}
 -\ft12 e^{-\hf-2(\a+\beta)\vf}\, {*G_{(2)\a}} \wedge G_{(2)}^\a \nn\\
&& -\ft12 e^{-(\hf+4\beta\vf)}\, {*G_{(1)\a\beta}} \wedge
G_{(1)}^{\a\beta}
-\ft1{12} m^2 e^{-\hf+2(\a-3\beta)\vf }
{*G_{(0)\a\beta\gamma}}\,  G_{(0)}^{\a\beta\gamma} \ ,
\eea
and the reduction of the Yang-Kinetic term gives
\bea {\cal L}_2 &=& -\ft12 e^{-(\hf+4\vf)/2}\, {*
\left(F^I-\cF^\a\, \phi_\a^I\right)}\wedge
\left(F^I-\cF^\beta\,\phi_\beta^I\right)
-\ft12 e^{-(\hf+4\beta\vf)/2}\, \, {*P^I_\beta} \wedge P^{I\beta} \nn\\
&& -\ft14 e^{-\ft12\hf +2(\a-2\beta)\vf}\, {*F^I_{\a\beta}}\,  F^{I\a\beta}\ .
\eea
It is important to note that any raising of the
group manifold indices from their original positions is performed with
$h^{\a\beta}$, which is the always well-defined inverse of
the unimodular scalar matrix $h_{\a\beta}$.
Thus, for example,
\be
F^{I\alpha\beta} \equiv h^{\alpha\gamma}\, h^{\beta\delta}\,
F^I_{\gamma\delta}\,.
\ee
Both the results given above are up to the group manifold volume
factor. In order to identify  the factor multiplying the $3$-form
kinetic term as the dilaton of the reduced theory, and to have
canonically-normalized scalar kinetic terms, we need to define the
scalar fields in $d$ dimensions as
\be
\phi = a^{-1} (\ah\,{\widehat\f}-4\a\,\vf)\,,\qquad
\sigma = a^{-1}(\ah\,\vf +4\a\,{\widehat\f})\,,
\ee
with
\be a^2= {8\over d-2}\,. \ee
These imply
\be \varphi = a^{-1} (\ah\,\sigma-4\a\,\phi)\,,
\qquad {\widehat\phi}= a^{-1}(\ah\,\phi +4\a\,\sigma)\,. \label{sd}
\ee
With these definitions at hand, the sum of ${\cal L}_3$, ${\cal L}_2$, the
Einstein-Hilbert term and the kinetic terms for ${\widehat \phi}$
produces the total $d$-dimensional bosonic Lagrangian
\bea {\cal L}_B  &=& R\, {*\oneone} - \ft12 {*d\phi}\wedge d\phi
-\ft12 e^{a\,\phi }\, {*G_\3}\wedge G_\3 \,
\nn\\
&& - \ft12{*d\sigma}\wedge d\sigma- {*P^{ij}}\wedge P^{ij} -\ft12 \,
{*P^{iI}}\wedge P^{iI}-\ft12 {*G_{\1}^{ij}}\wedge G_{\1}^{ij} \nn\w2
&&- \ft12 e^{a\, \phi /2} \, {*\cF^i}\wedge \cF^i- \ft12 e^{ a\phi
/2}\,*G_{\2}^i\wedge G_{\2}^i
\nn\\
&&-\ft12 e^{a \phi/2}\,{* \left(F^I-\cF^\a\,\phi_\a^I\right)}\wedge
\left(F^I-\cF^\beta\,\phi_\beta^I\right) -V\,
{*\oneone}\,,\label{fulllag} \eea
where the potential  $V$ for the scalar fields is given by
\bea
V &=& \ft14 e^{-a \phi/2} \left( F^I_{ij}F^{Iij}
+\ft13 \,G_{ijk}\,G^{ijk}\right)\nn\w2
&& +\ft14\, e^{-(a\, \phi + b\, \sigma)/2}\, \left(h_{\a\beta}\,
h^{\gamma\delta}\, h^{\rho\sigma}\, f^\a{}_{\gamma\rho}\,
f^\beta{}_{\delta\sigma} + 2 h^{\a\beta}\, f^\gamma{}_{\delta\a}\,
f^\delta{}_{\gamma\beta}\right)\,. \eea
Note that we have made the definitions
\bea
\cF^i&=&\cF^\a\,L_\a^i e^{b\sigma/4}\,,
\quad\quad\qquad\quad F_{ij}^I=F_{\a\beta}^I\,L^\a_i L^\beta_j\,
e^{-b\sigma/2}\,, \nn\w2
G_{\2 i}&=&G_{\2\a}\,L^\a_i\,e^{-b\sigma/4}\ ,
\qquad \quad G_{\1 ij}=G_{\1\a\beta}\,L^\a_i L^\beta_j\,e^{-b\sigma/2}\nn\w2
G_{ijk}&=& G_{\a\beta\gamma}\,L^\a_i L^\beta_j L^\gamma_j\,e^{-3b\sigma/4}
\,,\qquad
P^{iI}=P_\a L^{\a i}\,e^{-b\s/4}\,,
\eea
where
\be
b=\sqrt{8\over n}\,.
\ee
The raising and lowering of the indices $i,j,..=1,...,n$ and
$I=1,..., \hbox{dim}\,K$ are performed with the Kronecker deltas
$\delta_{ij}$ and $\delta_{IJ}$.

It is also useful to write the potential as the sum of squares of
the functions that appear in the supersymmetry transformation rules.
The result is:
\be V=-\ft1{24} \left(G_{ijk}-3 e^{-b\sigma/4} C_{[i.jk]}\right)^2
+\ft18 \left(G_{ijk} -e^{-b\sigma/4} C_{i.jk}\right)^2 +\ft14
\left(F_{ij}^I\right)^2\,. \ee


\subsection{ The supersymmetry transformation rules}


The group reduction of supersymmetry transformations were first studied
in detail in \cite{Salam:1984ft}
in the context of $SU(2)$ reduction of the eleven dimensional supergravity.
We refer to that work for some details, such as the compensating
gauge transformations
that play a significant role. Here we shall simply give our results.
One improvement is
that not only are our results valid for the group reduction of
supergravities in arbitrary
dimensions, but we also settle fully the problem of diagonalisation of all
the fermionic
kinetic terms. In \cite{Salam:1984ft} and many other works that followed,
typically only a partial diagonalisation of the gravitino kinetic term
was performed.

In order to obtain the supersymmetry transformation rules for the
vielbein and dilaton, and the kinetic terms for all the fermions in canonical
diagonalized form,
we find that we need
to define the $d$-dimensional fermions and supersymmetry
parameters as follows:
\bea
\psi_a &=& e^{\a\vf/2}\left({\widehat\psi}_a +
\ft{a^2}{8}\,\Gamma_a\Gamma^i{\widehat\psi}_i\right)\,,\nn\\
\psi_i &=& e^{\a\vf/2}\left({\widehat \psi}_i+\ft{\ah}{4}\,
\Gamma_i {\widehat\chi}\right)             \,,\nn\\
\chi &=& e^{\a\vf/2}\left( { -\ft{\ah}{a}\,{\widehat\chi}+
\ft{a}{2}\,\Gamma^i {\widehat \psi}_i }\right)\,,\nn\\
\lambda &=& e^{a\phi/2}\,{\widehat\lambda}\,,\nn\\
\epsilon &=& e^{-\a\vf/2} {\hat\epsilon}\,.
\eea
The inverse relations are also useful:
\bea{\hat\psi}_a &=& e^{-\a\vf/2}\left(\psi_a -\ft18{{\ah}^2}\,
\Gamma_a\Gamma^i \psi_i- \ft1{32}na{\ah}^2\,\Gamma_a \chi\right)\,,\nn\w2
{\hat\psi}_i &=& e^{-\a\vf/2}\left(\psi_i-\ft18 {\ah}^2
\Gamma_i\Gamma^j\psi_j +\ft{{\ah}^2}{4a}\Gamma_i\chi\right) \,,\nn\w2
{\hat\chi} &=& e^{-\a\vf/2}\left( -\ft{\ah}{a} \chi+\ft{\ah}{2}
\Gamma^i\psi_i\right)\,.
\eea
Note that $\Gamma_\mu$ and $\Gamma^i$ are both $32\times 32$ and have the
symmetry properties inherited from the $\Gamma$-matrices in $D=10$. In particular
they obey the Clifford algebra $\{\Gamma_\mu,\Gamma_\nu\}=2\eta_{\mu\nu}$,
$\{\Gamma_i,\Gamma_j\}=2\delta_{ij}$, and  $\{\Gamma_\mu,\Gamma_i\}=0$.
It is convenient to work with these matrices to exploit the familiar properties
they inherit from $D=10$ inheritance, and also to avoid clutter in notation.
If desired, it is an easy matter to express $\Gamma_\mu$ and $\Gamma^i$ as
direct product of suitable $\Gamma$ matrices of the  Clifford algebras in $d$
and $n$ dimensions.

With the above definitions of the fermionic fields in $d$
dimensions, their kinetic Lagrangian is completely diagonalized and
takes the form
\be {\cal L}= -{\bar\psi}_\mu \Gamma^{\mu\nu\rho} D_\nu\psi_\rho
-\ft12{\bar\chi}\Gamma^\mu D_\mu \chi  -{\bar\psi}_i\Gamma^\mu D_\mu
\psi_i-\ft12 {\bar\lambda}^I\Gamma^\mu D_\mu \lambda^I\ . \ee
Next, we compute the supersymmetry transformation rules. A
straightforward calculation yields for the bosons
\bea
\delta e_\mu^a &=& \ft12 {\bar\epsilon} \Gamma^a \psi_\mu\,,\w2
\delta\phi &=& \ft12 {\bar\epsilon}\chi\,,\w2
\delta B_{\mu\nu}&=& e^{-a\phi/2}
\left({\bar\epsilon}\Gamma_\mu\psi_\nu -\ft{a}{4}\,
{\bar\epsilon}\Gamma_{\mu\nu}\chi\right)
 -2 \delta A^\a_{[\mu}\, C_{\nu]\a}-A^I_{[\mu}\delta
A^I_{\nu]}\,, \w2
L_\a^i \delta \cA_\mu^\alpha &=&  e^{-(a\phi+b\sigma)/4}\left(
-\ft12{\bar\epsilon}\Gamma^i\psi_\mu -\ft{a}{8}\,{\bar\epsilon}
\Gamma_\mu\Gamma^i\chi -\ft12 {\bar\epsilon}\Gamma_\mu\psi^i
\right)\,,\w2
L^\a_i \delta C_{\mu\a} &=& e^{(b\sigma-a\phi)/4}
\left(-\ft12 {\bar\epsilon}\Gamma_i\psi_\mu -\ft{a}{8}\,
{\bar\epsilon}\Gamma_\mu\Gamma_i\chi +\ft12 {\bar\epsilon}\Gamma_\mu\psi_i \right)
\nn\\
&& +L^\a_i \left(-\delta \cA_\mu^\beta\,C_{\a\beta}+\delta A_\mu^I \phi_\a^I\right)
\,,\w2
\delta A_\mu^I &=& \ft12
e^{-a\phi/4}\,{\bar\epsilon}\Gamma_\mu\lambda^I +\delta
\cA_\mu^\a\,\phi_\a^I\,,\w2
\delta\sigma &=& -\ft{b}{4}\,{\bar\epsilon}\Gamma^i\psi_i\,, \nn\w2
L^\a_i L^\beta_j \delta B_{\a\beta} &=&
e^{b\sigma/2}\,{\bar\epsilon}\Gamma_{[i}\psi_{j]} -
\phi^I_{[i}\delta\phi^I_\beta\,L^\beta_{j]}\,, \w2
L^\a_i\delta L_{\a j} &=& \ft12 {\bar\epsilon}\Gamma_{(i}\psi_{j)}
-\ft1{2n}\delta_{ij}\,{\bar\epsilon} \Gamma^k\psi_k\,,\w2
L^\a_i \delta \phi_\a^I &=& \ft12
e^{b\sigma/4}\,{\bar\epsilon}\Gamma_i\lambda^I\,,
\eea
where $\phi_i^I=\phi_\a^I \,L^\a_i$.  Our results for the supersymmetry
transformations of the fermionic fields are:
\bea
%
\delta\psi_\mu&=& {\cal D}_\mu\epsilon +\ft1{96}\,e^{a\phi/2}\left(
a^2 \Gamma_\mu\Gamma^{\nu\rho\sigma}-12 \delta_\mu^\nu
\Gamma^{\rho\sigma}\right) \,G_{\nu\rho\sigma}\,\epsilon\nn\w2
&&  +\ft1{64}\,e^{a\phi/4}\left( a^2
\Gamma_\mu\Gamma^{\nu\rho}-16\delta_\mu^\nu\Gamma^\rho\right)
\left(\cF_{\mu\nu i} + G_{\mu\nu i}\right)\Gamma^i\,\epsilon\nn\w2
&& -\ft{a^2}{192}\,e^{-a\phi/4}\,  \left(G_{kij}-3 e^{-b\sigma/4}\,C_{k,\,ij}\right)\,\Gamma_\mu
\Gamma^{ijk}\epsilon \,,\w4
%
%
\delta \chi&=& \ft12 \Gamma^\mu \partial_\mu\f\, \e +\ft{a^2}{24}\,
e^{a\f/2}\,\Gamma^{\mu\nu\rho}G_{\mu\nu\rho}\,\epsilon +
\ft{a}{16}\,e^{a\phi/4}   \left( \cF_{\mu\nu i} + G_{\mu\nu
i}\right) \Gamma^{\mu\nu}\Gamma^i\,\epsilon \nn\w2
&& -\ft{a}{48}\,e^{-a\phi/4} \left(G_{kij}-
  3 e^{-b\sigma/4}\,C_{k,\,ij}\right)\,\Gamma^{ijk}\,\epsilon\,,\w4
\delta \psi_i &=& \ft18\,b\,\Gamma_i\Gamma^\mu\partial_\mu
\sigma\, \epsilon +\ft14 \left(G_{\mu ij}-2 P_{\mu ij}\right)\,
\Gamma^\mu\Gamma^j\,\epsilon
- \ft18\, e^{a\phi/4}\,\Gamma^{\mu\nu} \left( \cF_{\mu\nu i}
-G_{\mu\nu i} \right)\,\epsilon  \nn\w2
&& -\ft18\, e^{-a\phi/4} \left(
G_{ijk}-e^{-b\sigma/4}\,C_{i,jk}\right)\,\Gamma^{jk}\,\epsilon\,,\w4
%
%
\delta\lambda^I &=& -\ft14 e^{\a\phi/4}\,\left[\,\Gamma^{\mu\nu}\left(F^I_{\mu\nu}
-\cF_{\mu\nu}^\a\,\phi_\a^I\right)
+2\Gamma^\mu\Gamma^i P_\mu^{iI} +\Gamma^{ij}
F^I_{ij}\,\right]\epsilon\,, \eea
where we have used the value $\ah=-1$, and the modified covariant
derivative is given by
\be {\cal D}_\mu \epsilon = \partial_\mu \epsilon +\ft14
\omega_\mu{}^{ab}\,\Gamma_{ab}\,\epsilon  + \ft14\left( Q_{\mu ij}-\ft12
G_{\mu ij}\right)\,\Gamma^{ij}\epsilon\ .\ee
In particular, the minimal coupling of gauge fields in this covariant
derivative is given by
\be
{\cal D}\e= \partial_\mu\epsilon +
\ft14\,f^\gamma_{\a\beta}\,L^\a_i\left[ \left(g\,
L_{\gamma j}+ m L^\delta_j\,\eta_{\gamma\delta}\right)
A_\mu^\beta-g\,L^\beta_j
B_{\mu\gamma}\right]\Gamma^{ij}\,\epsilon +\cdots\label{cd}
\ee

Note also that the sum
$({\cF}+G_{\2})$ arises in the transformation rules of the graviton
multiplet, and as such these are the graviphotons, while the combination
$(\cF-G_{\2})$ occurs in the matter sector.


\section{The Hidden Duality Symmetries }


\subsection{The Bosonic Lagrangian}

The group reduction described above gives rise to half-maximal (i.e.
$16$ supersymmetries) gauged supergravities coupled to $n$ vector
multiplets in $d=(10-{\rm dim\ G})$ dimensions, where $n={\rm dim\
G}$. The scalar fields described by the internal metric
$h_{\a\beta}$ parametrize the coset $SL(n,R)/SO(n)$. The global
$SL(n,R)$ symmetry is manifest in the absence of gauging, but it
breaks down to $G\subset SL(n,R)$ upon gauging. Taking into account
the scalars $(B_{\a\beta}, \sigma, \phi_\a^I)$ as well, one expects
that together with $h_{\a\beta}$ they should parameterize the enlarged
coset $SO(n,n+{\rm dim\,K})/SO(n)\times SO(n+{\rm dim\,K})$. Indeed,
setting the gauge coupling constants to zero, the kinetic terms for
the vector fields take the form \cite{Maharana:1992my}
\be {\cal L}_{vec}= -\ft12 e^{-a\phi/2}\, {*(\cV^T\cG)^T }\wedge
\cV^T\cG\,, \label{vec}
\ee
where
\be \cG = \left(
           \begin{array}{c}
             d\cA^\a \\
             dB_\a \\
            dA^I \\
           \end{array}
         \right)
\ ,\qquad \cV \cV^T = M\,.
\ee
The matrix $M$ is given by
\bigskip
\be M=\left(
 \begin{array}{c|c|c}
    G+ \phi^T\phi +C^T G^{-1} C & C^T G^{-1} & -(1+C^T G^{-1})\phi^T \\
    \hline
    G^{-1} C& G^{-1} & -G^{-1}\phi^T \\
    \hline
   -\phi(1+ G^{-1}C)& -\phi G^{-1}& \oneone +\phi G^{-1}\phi^T \\
  \end{array}
\right)
\ee
\bigskip
where $G_{\a\beta}=\hL_\a^i\hL_\beta^i$ with the
definition ${\widehat L}_\a^i=e^{b\sigma/4} L_\a^i$, and
\be \cV=\left(
          \begin{array}{c|c|c}
            {\hL} & C^T \hL^{T-1} & -\phi^T \\
            \hline
            0 & \hL^{T-1} & 0 \\
            \hline
            0 & -\phi \hL^{T-1} & \oneone \\
          \end{array}
        \right)\,. \label{cV}
\ee
The inverse of this coset representative is given by
\be \cV^{-1}=\left(
          \begin{array}{c|c|c}
            \hL^{-1} & \hL^{-1}C & \hL^{-1}\phi^T \\
            \hline
            0 & \hL^T & 0 \\
            \hline
            0 & \phi & \oneone \\
          \end{array}
        \right)\,. \label{inverse}
\ee
The matrix $M$ leaves invariant the metric
\be
\eta = \left(
        \begin{array}{ccc}
          0 & -1 & 0 \\
          -1& 0& 0 \\
          0 & 0& 1 \\
        \end{array}
      \right)\,,   \label{ck2}
\ee
i.e. $M\eta M^T=\eta$, and therefore it forms the fundamental
representation of $SO(n,n+{\rm dim\,K})$. It also follows that
$\cV^{-1}=\eta \cV^T \eta$. The Lagrangian \eq{vec} is clearly
invariant under a global $SO(n,n+{\rm dim\,K})$. Turning on the gauge
couplings break this symmetry symmetry down to $K\times G \ltimes
R^n$.

The bosonic Lagrangian for  the scalars $(\sigma,
L_\a^i,B_{\a\beta})$ can be written as a sigma model on the coset
$SO(n,n+{\rm dim\,K})/SO(n)\times SO({\rm dim\,K})$. To this end we
define the symmetric Maurer-Cartan form,
\be
\cP=\ft12\left(\cV^{-1}d\cV+ (\cV^{-1}d\cV)^T\right)\nn\w2
= \left(
\begin{array}{c|c|c}
P_\mu^{ij}+\ft{b}4 \partial_\mu\sigma\delta^{ij}& 
  -\ft12G_{\mu ij} & -\ft12 P_\mu^{iI} \\
\hline
-\ft12 G_{\mu ij} &  
   -P_\mu^{ij}-\ft{b}4 \partial_\mu\sigma\delta^{ij}  & -\ft12 P_\mu^{iI}\\
\hline -\ft12 P_\mu^{iI} & -\ft12 P_\mu^{iI} & 0
\end{array}\right)\,.\label{pp}
\ee
With the aid of this formula, and recalling \eq{vec}, the full bosonic 
Lagrangian for
abelian gauge fields with no gauging can be put into the 
remarkably simple form
\be {\cal L}_B = R\, {*\oneone} - \ft12 {*d\phi}\wedge d\phi -\ft12
e^{a\,\phi }\, {*G_\3}\wedge G_\3 - \ft12 \tr {*\cP} \wedge \cP -\ft12
e^{-a\phi/2}\, {*(\cV^T\cG)^T }\wedge \cV^T\cG\,, \label{bL} \ee
where
\be G_\3=dB_2 -\ft12 \cG^\ua \wedge \cB_\ua\,,
\ee
with the underlined Greek indices understood  to be contracted by the
Cartan-Killing metric given in \eq{ck2}.\\

\subsection{The Supersymmetry Transformation Rules}

The $SO(n,n+{\rm dim}\,K)$ symmetry can be made manifest in the
supersymmetry transformation rules as well. To begin with, we combine
the vector fields and fermionic fields as
\be \cB_\mu^\ua = \left(
           \begin{array}{c}
             \cA_\mu^\a \\
             B_{\mu\a} \\
            A_\mu^I \\
           \end{array}
         \right)\ ,\qquad\qquad  \psi^r = \left(
           \begin{array}{c}
             \psi^i \\
            {1\over \sqrt 2}\,\lambda^I \\
           \end{array}
         \right)
\ee
Furthermore, we define
\be \cV=\left(\cV_\ua^i,\cV_\ua^r\right)\ ,\qquad i=1,...,n,\quad
r=1,...,(n+{\rm dim}\,K)\,, \ee
where
\be \cV_\ua^i=\cV_\ua^{+i}\ ,\qquad
\cV_\ua^r=\left(\cV_\ua^{-i}, \cV_\ua^I\right)\ ,\qquad \cV_\ua^{\pm
i}= {1\over \sqrt 2}\,\left(\pm\cV_\ua^{1i} + \cV_\ua^{2i}\right)\ .
\label{defs} \ee
Here, $\cV_\ua^{1i}$ and $\cV_\ua^{2i}$ represent the two 
$n$ by $(2n+{\rm dim}\, K)$
blocks in the matrix $\cV$ given in \eq{cV}.  Explicitly, we have
\be
(\cV_\ua^i, \cV_\ua^r)= \left(
          \begin{array}{c|c|c}
            \ft1{\sqrt 2} (\hL + C^T \hL^{T-1}) & 
                       \ft1{\sqrt 2} (-\hL + C^T \hL^{T-1})& -\phi^T \\
            \hline
            \ft1{\sqrt 2}\hL^{T-1} & \ft1{\sqrt 2}\hL^{T-1} & 0 \\
            \hline
            -\ft1{\sqrt 2}\phi \hL^{T-1} & 
                 -\ft1{\sqrt 2}\phi \hL^{T-1} & \oneone \\
          \end{array}
        \right)\,, \label{cVpm}
\ee
which has the inverse
\be (\cV^\ua_i , \cV^\ua_r)= \left(
          \begin{array}{c|c|c}
            \ft1{\sqrt 2} \hL^{-1} & \ft1{\sqrt 2} 
(\hL^{-1}C+\hL^T)& \ft1{\sqrt 2}\hL^{-1}\phi^T \\
            \hline
            -\ft1{\sqrt 2}\hL^{-1} & 
  \ft1{\sqrt 2}(-\hL^{-1}C+\hL^T )& -\ft1{\sqrt 2}\hL^{-1}\phi^T \\
            \hline
            0 & \phi & \oneone \\
          \end{array}
        \right)\,, \label{cVpmi}
\ee
These definitions are needed in order to ensure that the
combinations $(\cA_\mu + B_\mu)$, representing the graviphotons, and
$(\cA_\mu - B_\mu)$, which are the external vector fields, transform
appropriately in relation to the fermions of the supergravity and vector
multiplets respectively. Denoting the coset representative defined
above by $\cV_{\rm diag}$ and its inverse by $\cV_{\rm diag}^{-1}$,
they satisfy the relation $\cV_{\rm diag}^{-1}= \eta_{\rm diag}\,
\cV^T_{\rm diag} \, \eta$ where
\be \eta_{\rm diag } = \left(
        \begin{array}{ccc}
          -1 & 0 & 0 \\
          0& 1& 0 \\
          0 & 0& 1 \\
        \end{array}
      \right)\,,   \label{diag}
\ee
Armed with these definitions, we can now show that the supersymmetry
transformations of all the bosonic fields simplify dramatically, and
they can be written in a manifestly $SO(n,n+{\rm dim\,K})$ covariant
way, as follows:
\bea
\delta e_\mu^a &=& \ft12 {\bar\epsilon} \Gamma^a \psi_\mu\,,\label{hs1}\w2
\delta\phi &=& \ft12 {\bar\epsilon}\chi\,,\w2
\delta B_{\mu\nu}&=& e^{-a\phi/2}
({\bar\epsilon}\Gamma_{[\mu}\psi_{\nu]} -\ft{a}{4}\,
{\bar\epsilon}\Gamma_{\mu\nu}\chi)
 - \delta \cB^\ua_{[\mu}\, \cB_{\nu]\ua}\,, \w2
\delta\cB_\mu^\ua &=& -\ft1{\sqrt 2}
e^{-a\phi/4}({\bar\epsilon}\Gamma^i\psi_\mu+\ft{a}{4}\,
{\bar\epsilon}\Gamma_\mu\Gamma^i\chi)\,\cV^\ua_i
+\ft1{\sqrt 2} e^{-a\phi/4}\,{\bar\epsilon}\psi^r\,\cV^\ua_r\,,\w2
\left(\cV^{-1}\delta \cV\right)_{ir} &=& -\ft14 {\bar\epsilon}\Gamma_i\psi_r\ .\label{hs5}
\eea
In the last equation, bearing in mind \eq{pp} and \eq{defs}, it is
understood that $\left(\cV^{-1}\delta
\cV\right)_{ij}=-\ft12\left((L^{-1}\delta L)_{ij} +\ft{b}4
\delta\sigma \delta^{ij}\right) -\ft14(L^{-1}\delta B L^{-1})_{ij}$
and that $\left(\cV^{-1}\delta \cV\right)_{iI}= -\ft12 (L^{-1}\delta
\phi^T)_{iI}$.

The supersymmetry transformations of the fermionic fields also 
simplify considerably
in a manifestly duality-symmetric form as follows:
\bea
%
\delta\psi_\mu&=& {\cal D}_\mu\epsilon +\ft1{96}\,e^{a\phi/2}\left(
a^2 \Gamma_\mu\Gamma^{\nu\rho\sigma}-12 \delta_\mu^\nu
\Gamma^{\rho\sigma}\right) \,G_{\nu\rho\sigma}\,\epsilon\nn\w2
&&  +\ft1{32}\,e^{a\phi/4}\left( a^2
\Gamma_\mu\Gamma^{\nu\rho}-16\delta_\mu^\nu\Gamma^\rho\right)
\cG_{\mu\nu}^i \Gamma_i\,\epsilon\,,\w4
%
%
\delta \chi&=& \ft12 \Gamma^\mu \partial_\mu\f\, \e +\ft{a^2}{24}\,
e^{a\f/2}\,\Gamma^{\mu\nu\rho}G_{\mu\nu\rho}\,\epsilon +
\ft{a}{8}\,e^{a\phi/4} \cG_{\mu\nu}^i
\Gamma^{\mu\nu}\Gamma^i\,\epsilon\,,\w4
\delta \psi^r &=& \cP_\mu^{ir} \Gamma^\mu\Gamma^i\epsilon
-\ft14 e^{a\phi/4} \cG_{\mu\nu}^r\Gamma^{\mu\nu}\epsilon\,,
\eea
where $\cG^i=\cG^\ua\,\cV_\ua^i$ and $\cG^r=\cG^\ua\,\cV_\ua^r$. 
The covariant derivative
of the supersymmetry parameter reads
\be
 {\cal D}_\mu \epsilon = \partial_\mu \epsilon +\ft14
\omega_\mu{}^{ab}\,\Gamma_{ab}\,\epsilon  + \ft14 \cQ_{\mu
ij}\,\Gamma^{ij}\epsilon\ , \label{cd1}\ee
with the $SO(n)$-valued composite connection given by
$\cQ_{\mu ij}= \cV^{-1}_{[i}\partial_\mu \cV_{j]}$.
Here we have used the definitions given in \eq{defs}.

\subsection{The nonabelian model}

The hidden $SO(n,n+{\rm dim}\,K)/SO(n)\times SO(n+{\rm dim}\,K)$
coset structure can be maintained in the gauged model with
nonabelian gauge symmetry as well. To show this, we begin with the
observation that the nonabelian field strength
\be \cG^\ua= d\cB^\a +\ft12 {\hat f}^\ua{}_{\ub\uc}\ \cB^\ub \wedge
\cB^\uc\ ,\quad\quad  \ua=\a,\a',I\ , \ee
produces the three field strengths \eq{mfs}, \eq{efs} and \eq{g2}
with the only nonvanishing structure constants given by
\be \fh^\a{}_{\beta\gamma}= f^\a{}_{\beta\gamma}\ ,\   \quad
\fh^{\a'}{}_{\beta\gamma}= m\,f_{\a\beta\gamma}\ ,\quad
\fh^{\a'}{}_{\beta\gamma'}=
 -\fh^{\a'}{}_{\gamma'\beta}= 
-f^\gamma{}_{\beta\a}\ ,\quad \fh^I{}_{JK}= f^I{}_{JK}\ , \ee
These define the semi-direct-product group $K\times G\ltimes R^n$, and we
have the associated field strengths
\be \cG^\ua \cV_\ua = \left(
           \begin{array}{c}
             \cF^\a \\
             G_{\2\a} \\
           F^I-\cF^\ua \phi_\a^I \\
           \end{array}
         \right)
\ . \ee
With these identifications, we find that the field strength $G_\3$
defined in \eq{g3} can also be written in a manifestly $K\times
G\ltimes R^n$ invariant form as
\be G_\3 = dB_2 -\ft12 \left(\cG^\ua \wedge \cB_\ua -\ft16
\fh^\ua{}_{\ub\uc}\, \cB^\ub \wedge \cB^\uc \wedge \cB_\ua\right)\ ,
\ee
where $\cB_\ua= \cB^\ub \eta_{\ua\ub}$. Next, we turn to the
supersymmetry transformation rules.

The supersymmetry transformations of the bosonic fields will be as
given in \eq{hs1} -\eq{hs5}, while those of the fermionic fields
will now take the form
\bea
%
\delta\psi_\mu&=& {\cal D}_\mu\epsilon +\ft1{96}\,e^{a\phi/2}\left(
a^2 \Gamma_\mu\Gamma^{\nu\rho\sigma}-12 \delta_\mu^\nu
\Gamma^{\rho\sigma}\right) \,G_{\nu\rho\sigma}\,\epsilon\nn\w2
&&  +\ft1{32}\,e^{a\phi/4}\left( a^2
\Gamma_\mu\Gamma^{\nu\rho}-16\delta_\mu^\nu\Gamma^\rho\right)
\cG_{\mu\nu}^i \Gamma_i\,\epsilon -\ft{a^2}{48\sqrt 2} \Gamma_\mu
T_{ijk}\Gamma^{ijk}\epsilon\,,\w4
%
%
\delta \chi&=& \ft12 \Gamma^\mu \partial_\mu\f\, \e +\ft{a^2}{24}\,
e^{a\f/2}\,\Gamma^{\mu\nu\rho}G_{\mu\nu\rho}\,\epsilon +
\ft{a}{8}\,e^{a\phi/4} \cG_{\mu\nu}^i
\Gamma^{\mu\nu}\Gamma^i\,\epsilon -\ft{a}{12\sqrt 2}
T_{ijk}\Gamma^{ijk}\epsilon\,,\w4
\delta \psi^r &=& \cP_\mu^{ir} \Gamma^\mu\Gamma^i\epsilon -\ft14
e^{a\phi/4} \cG_{\mu\nu}^r\Gamma^{\mu\nu}\epsilon -\ft1{2\sqrt 2}
T_{ij}^r \Gamma^{ij}\epsilon\,, \eea
where the covariant derivative $D_\mu\epsilon$ is defined as in
\eq{cd1}, with the composite connection now given by
$\cQ_{ij}=\cV^{-1}_{[i}D_\mu \cV_{j]}$, involving the $K\times
G\ltimes R^n$ covariant derivative.

Finally, we find that the bosonic action can also be written in a
manifestly $K\times G\ltimes R^n$ invariant form as
\bea {\cal L}_B &=& R\, {*\oneone} - \ft12 {*d\phi}\wedge d\phi
-\ft12 e^{a\,\phi }\, {*G_\3}\wedge G_\3- \ft12 \tr {*\cP} \wedge \cP
\nn\w2
&&  -\ft12 e^{-a\phi/2}\, {*(\cV^T\cG)^T }\wedge \cV^T\cG -\ft13
T_{ijk}T^{ijk} + T_{ijr} T^{ijr}\ , \label{bnL}\eea
where
\be \cP_\mu= \cV^\ub\left( \partial_\mu\delta_\ub^\ua +
\fh^\ua{}_{\ub\uc}\,\cB_\mu^\uc\right) \cV_\ua\ ,\ee
and the $T$-tensors are just the boosted structure constants, defined
by
\be
T_{ijk}= {\hat f}^\ua{}_{\ub\uc}\,\cV_{\ua i}\,\cV^\ub_j\,\cV^\uc_k\ ,
\quad\quad
T_{ijr}= {\hat f}^\ua{}_{\ub\uc}\,\cV_{\ua r}\,\cV^\ub_j\,\cV^\uc_k\ .
\ee
The sigma model kinetic term involves $\cP_\mu$, which takes the same
form as in \eq{pp} but with covariantized field strengths.

\section{ The Structure and Consistent Truncations}

The results we obtained above have the same overall structure as in the
couplings of half-maximal $d$-dimensional supergravities coupled
to $(n+{\rm dim}\,K)$ vector multiplets. However, while in a typical
gauged supergravity theory it is usually assumed that the gauge
group is semisimple, here we have landed on a particular class of
theories in which the gauge group is $K\times G\ltimes R^n$, where
$K$ is semisimple but $G$ is a completely arbitrary
group of dimension $n <10$ with traceless structure constants. The semi-direct
structure brings in restrictions on the consistent truncations of
the vector multiplets, as we shall see in the next section. In
particular, looking at a case of particular interest, namely the
$d=7$ theory with gauge symmetry $SO(2,1)\ltimes R^3$ (with the
gauge fields of group $K$ set to zero), we find that the two vector
fields that need to be truncated away in order to obtain the standard
$SO(2,1)$ gauged model of \cite{Bergshoeff:2005pq} do not actually
truncate consistently. This is due to the fact that the six vector
fields in the theory split into a triplet in the adjoint
representation of $SO(2,1)$, with the remaining (external) triplet of
vector fields transforming nontrivially under $SO(2,1)$.

On the other hand, a consistent truncation of the external $n+
{\dim}\,K)$ vector multiplets in which $n$ vector multiplets that
gauge the compact $R$-symmetry group always
exists, as we shall see below.  We shall first describe this 
successful truncation, and after that, 
show explicitly the  obstacles one faces in attempting to truncate the
theory in $d=6$ to a gauged chiral supergravity.  This is closely
related to the consistent truncation problem mentioned above.
Indeed, had the standard $SO(2,1)$ gauged theory in $d=7$ resulted
from our model, it is known \cite{Bergshoeff:2005pq} that it could then
by reduction and truncation produce a chiral gauged supergravity in 
six dimensions.

\subsection{The truncation of the external vector multiplets}

In $d$ dimensions, the model we have obtained describes the gauged
coupling of  half-maximal supergravity,
\be (g_{\mu\nu}, \
B_{\mu\nu}, \ \phi, \ \cA_\mu^i+B_\mu^i,\  \psi_\mu,\ \chi)\ , \ee
with $8(d-2)_B+8(d-2)_F$ degrees of freedom, coupled to $n+{\rm
dim\,G}$ vector multiplets with field content
\be (\cA_\mu^i -B_\mu^i,\ h_{\a\beta}, \ B_{\a\beta}, \ \sigma, \
\psi_i)\ ,\qquad (A_\mu^I, \lambda^I, \phi_\a^I)\,,
\ee
and with $8(n+{\rm dim\,G})_B + 8(n+{\rm dim\,G})_F$ degrees of freedom.
All the vector multiplets can be consistently decoupled, to leave a
gauged half-maximal supergravity with gauge group $G$. To see this,
we set
\bea
 h_{\a\beta}&=&h_{\a\beta}^\0\,,\qquad  \cA_\mu^i-B_\mu^i=0\,,
\qquad A_\mu^I=0\,,\nn\w2
\sigma&=&0\,, \qquad \ B_{\a\beta}=0\,,\qquad  \qquad \phi_\a^I=0\,,\nn\w2
\psi_i &=& 0\ , \qquad \lambda^I=0\,, \eea
where $h_{\a\beta}^\0$ is a constant unimodular $4\times 4$ matrix.
With this ansatz, the supersymmetry transformation rules consistently
truncate, provided that
\be
\cF_{\mu\nu}^i - G_{\mu\nu}^i=0\,, \ \ G_{ijk}-C_{k,\,ij}=0\,,\ \
G_{\mu ij}=0\,,\ \  P_{\mu ij}=0\,, \ \sigma=0\,.
\ee
The first and second conditions independently give the relation
\be
m\,\eta_{\gamma\delta}\,f^\delta{}_{\a\beta}
+ g\left(2f^\delta{}_{\gamma[\a}\,h^\0_{\beta]\delta} -f^\delta{}_{\a\beta}\,
h^\0_{\gamma\delta}\right)=0\,,\label{r1}
\ee
while the third and fourth conditions, respectively, give the relations
\be
 f^\delta{}_{\a\beta}\left(g\,h^\0_{\gamma\delta}+
  m\eta_{\gamma\delta}\right)=0\,,
 \qquad g\,f^\delta{}_{\gamma(\a}\,h^\0_{\beta)\delta}=0\,.\label{r2}
\ee
All of the above conditions are satisfied by taking
\be h_{\a\beta}^\0=\eta_{\a\beta}=\delta_{\a\beta}\,,
\qquad g=-m\,.\label{sol}
\ee
Finally, it is important to ensure that all  field equations for the
fields that we have eliminated are also satisfied, since not all of
them may follow from the integrability conditions of the local
supersymmetry transformations. We have checked that indeed all the
required field equations are satisfied by the solution \eq{sol}.
Thus, the resulting supergravity is indeed gauged half-maximal
supergravity with $8(d-2)_B+8(d-2)_F$ degrees of freedom, with a type A
gauge group $G$ associated with the gauge fields
$(A_\mu^i+B_\mu^i)$.

\subsection{The nonexistence of chiral truncation in $d=6$}

At first sight, the most general chiral truncation would require setting
\be \psi_{\mu -}=0\ ,\ \ \chi_+=0\ ,
\ee
leaving us with the $N=(1,0)$ supergravity and tensor multiplet, 
which contains the fields
$(g_{\mu\nu},\ B_{\mu\nu},\ \phi,\ \psi_{\mu +},\ \chi_-)$,
the vector multiplets
$( \cA_\mu^i-B_\mu^i,\ \psi_{i+} )$ and
$(A_\mu^I,\ \lambda_+^I)$, and the hypermultiplets
$(h_{\a\beta},\ B_{\a\beta},\ \sigma,\ \psi_{i-})$ and
$(\phi_\a^I,\ \lambda_-^I)$.
However, this approach immediately runs into trouble, as can be seen
by examining the
$\delta\psi_{\mu -}=0$ condition, since this gives
\be
\cF_{\mu\nu}^i+G_{\mu\nu}^i=0\,,
\qquad G_{ijk}-3e^{-b\sigma/4}\,C_{[k,\,ij]}=0\,. \label{ch1}
\ee
These equations place constraints on the fields of the vector and
hypermultiplets listed above. The second condition, for example,
imposes a nonlinear algebraic constraint on the scalar fields
$(\sigma, h_{\a\beta}, \phi_\a^I)$, and it is not clear at all if
this can be satisfied with any surviving hypermultiplet. This
condition alone motivates us to eliminate all the hypermultiplets.
Thus, in particular setting $\sigma=0$, $h_{\a\beta}=h_{\a\beta}^\0$
with constant $h_{\a\beta}^\0$, and $\psi_{i -}=0$, we find from the
requirement $\delta\psi_{i -}=0$ that
\be
P_{\mu ij}=0\ ,\qquad G_{\mu ij}=0\,. \label{ch2}
\ee
These constraints,
and the second one in \eq{ch1} respectively, give the conditions
\bea
&&g \,f^\delta{}_{\gamma(\a}\,h^\0_{\beta)\delta}=0\,,\label{f1}\w2
&& f^\delta{}_{\a\beta}\left(g\,h^\0_{\gamma\delta}+
  m\eta_{\gamma\delta}\right)=0\,,
\label{f2}\w2
&& m\,\eta_{\a\delta}\,f^\delta{}_{\a\beta}-
  3g\,f^\delta{}_{[\a\beta}\,h_{\gamma]\delta}^\0=0\,.\label{f3}\eea
Recalling the requirement of $R$-symmetry gauging, by which the
minimal coupling in \eq{cd} must be nonvanishing,  we find that the
first equation above can only be satisfied for $SU(2)\times U(1)$,
in which case the remaining two equations clearly cannot be
satisfied. Thus, we conclude that the model
obtained by group manifold reduction cannot be consistently
truncated to yield a gauged chiral $N=(1,0)$ supergravity in $d=6$.


\section{Conclusions}


The heterotic supergravity with gauge symmetry $K$ reduced on a
group manifold $G$ of dimension $n$ gives rise to a large class of
gauge supergravities in $d=(10-n)$ dimensions that describe the coupling
of $(n+ {\rm dim}\,K)$ vector multiplets. The reduction involves a
nonvanishing $3$-form flux, and the Lie algebra of $G$ must have
traceless structure constants in order to ensure the consistency of the
reduction at the level of the action. The $d$-dimensional theory has
$K\times G\times R^n$ gauge symmetry, and its couplings are governed
by the coset $SO(n,n+{\rm}\, K)/SO(n)\times SO(n+{\rm}\, K)$, which 
is parametrized by all the scalars of the theory save the dilaton.  The
group $ SO(n,n+{\rm}\, K)$ is a global duality symmetry group that
is broken down to $SO(n)\times SO(n+{\rm}\, K)$ in the presence of
gauge couplings. A linear combination of the $n$ Kaluza-Klein
vectors and $n$ ``winding'' vectors coming from the $2$-form potential 
assembles into the
adjoint representation of the group $G$, while the orthogonal
combination transforms as matter vectors under this group. These
matter vectors and the gauge field of group $G$ can be consistently 
truncated in any dimension, to yield pure half-maximal gauged supergravity 
with a compact $R$-symmetry group. Truncations to yield noncompact
gaugings do not seem to be possible.

These results primarily provide an embedding of a large class of
half-maximal gauged supergravities in heterotic supergravity, and hence
in string theory. They also highlight the absence of such origins for
some noncompact gauged supergravities that are of considerable interest in
fewer than ten dimensions.

In this paper we have restricted our attention to Lie algebras with 
traceless structure
constants, since these give consistent reductions not only at the level of
the field equations, but also at the level of the action. 
The traceless algebras are referred to as type A Lie algebras. 
Reduction on group
manifolds based on type B algebras, for which the structure
constants have non-vanishing traces $f^\alpha{}_{\alpha\beta}\ne 0$,
is consistent only at the level of field equations.  Furthermore, the
resulting $d$-dimensional equations are not derivable from an
action. It would be useful to perform the reduction for type B
algebras as well, which is a straightforward matter using the
techniques of this paper, in order to see whether our conclusions about the
further consistent truncations persist. It would also be worthwhile to
explore the consequences of the free parameters that arise in a
number of Lie algebras occurring in the group reduction.

\bigskip\bigskip

{\bf Acknowledgments}

We thank Eric Bergshoeff, Murat G\"{u}naydin  and Kelly Stelle for
useful discussions. The work of H. L\"u and C.N.P. is supported in 
part by DOE grant DE-FG03-95ER40917, and that of E.S. is supported in 
part by NSF Grant PHY-0555575.

\appendix

\section{Lie algebras of dimension 2, 3 and 4}

\begin{table}
\begin{tabular}{|c|l|l|l|}\hline\hline
Name & Commutator & Type & $\eta_{\mu\nu}$\\ \hline\hline
$A_{2,1}$ & $e_{12}=e_2$ &B& $(-\ft12,0)$\\ \hline\hline
$A_{3,1}$ & $e_{23}=e_1$ &A& $(0,0,-1)$\\ \hline
$A_{3,2}$ & $e_{13}=e_1,e_{23}=e_1+e_2$ &B& $(0,0,-1)$\\ \hline
$A_{3,3}$ & $e_{13}=e_1,e_{23}=e_2$ &B& $(0,0,-1)$\\ \hline
$A_{3,4}$ & $e_{13}=e_1,e_{23}=-e_2$ &A& $(0,0,-1)$\\ \hline
$A^a_{3,5}$ & $e_{13}=e_1,e_{23}=ae_2\,, (0<|a|<1)$
              &B& $(0,0,-\ft12(1+a^2)$\\ \hline
$A_{3,6}$ & $e_{13}=-e_2,e_{23}=e_1$ &A& $(0,0,1)$\\ \hline
$A^a_{3,7}$ & $e_{13}=ae_1 - e_2,e_{23}=e_1 + a e_2\,,(a>0)$
                    &B& $(0,0,1-a^2)$ \\ \hline
$A_{3,8}$ & $e_{13}=-2e_2,e_{12}=e_1, e_{23}=e_3$
                    &A& $\{-2,-1,-2\}$\\ \hline
$A_{3,9}$ & $e_{13}=e_3,e_{23}=e_1, e_{31}=e_2$
                    &A& $(1,1,1)$\\ \hline\hline
$A_{4,1}$ & $e_{24}=e_1, e_{34}=e_2$ &A& $(0,0,0,0)$\\ \hline
$A^a_{4,2}$ & $e_{24}=ae_1, e_{24}=e_2, e_{34}=e_2+e_3, (a\ne 0)$
                    &A$(a=-2)$& $(0,0,0,-\ft{2-a^2}{2})$\\ \hline
$A_{4,3}$ & $e_{14}=e_1, e_{34}=e_2$ &B& $(0,0,0,-\ft12)$\\ \hline
$A_{4,4}$ & $e_{14}=e_1, e_{24}=e_1+e_2, e_{34}=e_2+e_3$
                    &B& $(0,0,0,-\ft32)$\\ \hline
$A^{ab}_{4,5}$ & $e_{14}=e_1, e_{24}=ae_2, e_{34}=be_3$ & & \\
            &$(ab\ne 0, -1\le a\le b\le 1)$ &A$(a+b=-1)$&
                $(0,0,0,-\ft{a^2+b^2+1}{2})$ \\ \hline
$A^{ab}_{4,6}$ & $e_{14}=ae_1, e_{24}=be_2-e_3,
          e_{34}=e_2+be_3$ & & \\
 & $(a\ne 0, b\ge 0)$ &A$(a=-2b)$ & $(0,0,0,\ft{2-a^2-2b^2}{2})$
                              \\ \hline
$A_{4,7}$ & $e_{23}=e_1, e_{14}=2e_2, e_{23}=e_2,e_{34}=e_2+e_3$
                      &B& $(0,0,0,-3)$\\ \hline
$A_{4,8}$ & $e_{23}=e_1, e_{24}=e_2, e_{34}=-e_3$
                     &A& $(0,0,0,-1)$\\ \hline
$A^b_{4,9}$ & $e_{23}=e_1, e_{14}=(1+b)e_2,
   e_{24}=e_2$ && \\
    & $e_{34}=be_3\,,(-1<b\le 1)$ &B &
           $(0,0,0,{\scriptstyle-1-b-b^2})$\\ \hline
$A_{4,10}$ & $e_{23}=e_1, e_{24}=-e_3, e_{34}=e_2$ &A&
       $(0,0,0,1)$\\ \hline
$A_{4,11}$ & $e_{23}=e_1, e_{14}=2ae_2, e_{24}=ae_2-e_3$ && \\
 & $e_{34}=e_2+ae_3\,, (a>0)$ &B& $(0,0,0,1-3a^2)$\\ \hline
$A_{4,12}$ & $e_{13}=e_1, e_{23}=e_2, e_{14}=-e_2,e_{24}=e_1$
&B& $(0,0,-1,1)$\\ \hline
\end{tabular}
\caption{\small The complete list of Lie algebras in dimensions up
to four that are not direct sums of lower dimensional ones. The
commutator $[e_i,e_j]$ is denoted by $e_{ij}$ for short. The last
column gives the Cartan-Killing metric which is diagonal except for
$A_{3,8}$ which is minor-diagonal shown by $\{ \}$ and represents
$SO(2,1)$.}
\end{table}

\newpage


\end{document}